%% file: main.tex
\documentclass[twocolumn]{article}

\usepackage[nolist]{acronym}
\usepackage{booktabs} 
\usepackage{subcaption}
\usepackage{mathtools, algorithm}
\usepackage{algorithmicx}  
\usepackage{algpseudocode}
\usepackage{multirow}
\usepackage{colortbl}
\usepackage{xcolor}
\usepackage{enumitem}
\setlist{nosep}

\input{math_commands}

\usepackage{microtype}
\usepackage{graphicx}
\usepackage[margin=0.8in]{geometry}
\usepackage{soul}

\usepackage{mathtools, algorithm}
\usepackage{algorithmicx}  
\usepackage{algpseudocode}

\usepackage{hyperref}
\usepackage[numbers]{natbib}
\PassOptionsToPackage{sort}{natbib}

\definecolor{OliveGreen}{rgb}{0,0.4,0}

\newcommand{\second}[1]{{\color{blue}{#1}}}
\newcommand{\green}[1]{{\color{OliveGreen}{#1}}}
\newcommand{\appendixhead}%
{\begin{center}
\textbf{\huge Appendix}
\end{center}}

\newcommand\blfootnote[1]{%
  \begingroup
  \renewcommand\thefootnote{}\footnote{#1}%
  \addtocounter{footnote}{-1}%
  \endgroup
}

\begin{document}
\title{Density Weighting for Multi-Interest Personalized Recommendation}

\author{
Nikhil Mehta$^*$, Anima Singh, Xinyang Yi, Sagar Jain, Lichan Hong and Ed H. Chi\\
Google
}
\graphicspath{{Figures/}}

\maketitle
\begin{abstract}
Using \ac{MUR} to model user behavior instead of a \ac{SUR} has been shown to improve personalization in recommendation systems. However, the performance gains observed with \ac{MUR} can be sensitive to the skewness in the item and/or user interest distribution. When the data distribution is highly skewed, the gains observed by learning multiple representations diminish since the model dominates on head items/interests, leading to poor performance on tail items. Robustness to data sparsity is therefore essential for \ac{MUR}-based approaches to achieve good performance for recommendations. Yet, research in \ac{MUR} and data imbalance have largely been done independently. In this paper, we delve deeper into the shortcomings of \ac{MUR} inferred from imbalanced data distributions. We make several contributions: (1) Using synthetic datasets, we demonstrate the sensitivity of \ac{MUR} with respect to data imbalance, (2) To improve \ac{MUR} for tail items, we propose an iterative density weighting scheme (IDW) with user tower calibration to mitigate the effect of training over long-tail distribution on personalization, and (3) Through extensive experiments on three real-world benchmarks, we demonstrate IDW outperforms other alternatives that address data imbalance.
\end{abstract}

\vspace{-0.5em}
\section{Introduction}
\blfootnote{$^*$This work was done when the author was an intern at Google. Correspondence to \{nikhilmehta,animasingh\}@google.com}
Recommender systems are at the core of various online services~\cite{yi2019samplingbias, Hallinan2016Netflix, Linden2003Amazon, Messica2018Ebay, abelTwitter, Pal2020Pinterest}, allowing users to discover items that are of potential interest from a large catalog. Therefore, it is vital for recommendation systems to model user interest behavior effectively. This can be especially challenging for sequential recommender system that takes a sequence of user's past behavior to predict user's future interaction, since user's behavior sequence typically reflects a diverse and dynamic interest profile.

\begin{figure}[t]
    \centering
    \includegraphics[width=0.95\linewidth]{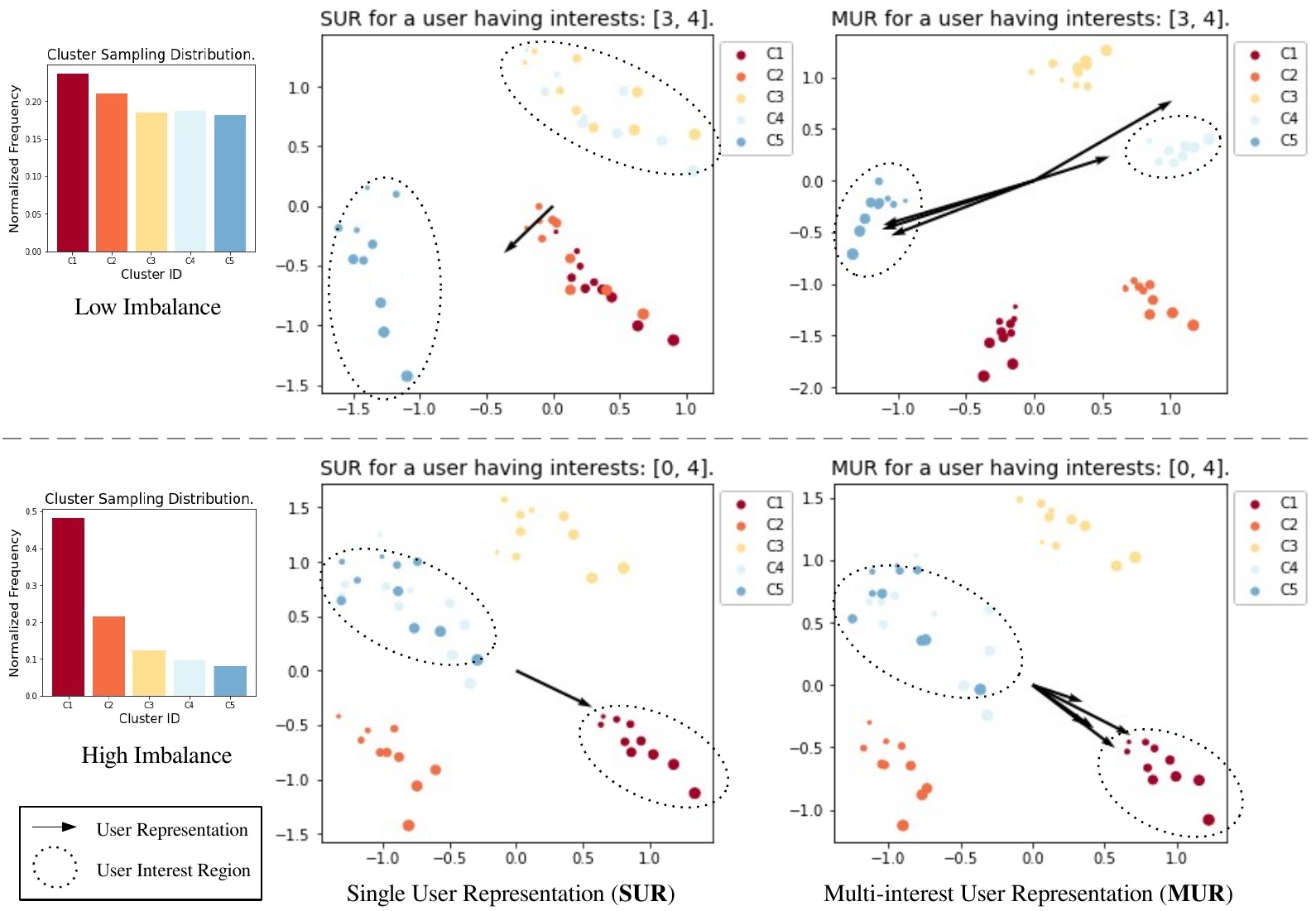}
    \caption{The learned item representations using Single (SUR) vs. Multiple User Representations (MUR) on synthetic user-item datasets with different levels of imbalance. The cluster-ids indicate various user interests. Top-row: Low imbalance in user interests. Bottom-row: High imbalance in user interests. The plots depict the learned item representations with SUR/MUR model, and the user representation learned for a random user with two interests.}
    \label{fig:sur-vs-mur}
    \vspace{-1.5em}
\end{figure}

A large scale recommendation system usually consists of two stages, the retrieval stage and the ranking stage. The retrieval stage is critical in narrowing down the candidates items that are relevant to user interests. Existing approaches formulate the retrieval task by assuming a unified low-dimensional representation space for users and items, where a user representation encodes the interest profile of the user and is used to retrieve top candidate items. Recent work~\cite{Li2019Tmall, Cen2020ControllableMF} has shown that a fixed-length representation can become a bottleneck in modeling user interests for large-scale datasets. Modern approaches~\cite{Cen2020ControllableMF, Kula2017MixtureoftastesMF} circumvent the limitations of a \acf{SUR} by using \acf{MUR} to model the holistic interest profile from the past behaviors of users. A synthetic experiment in Figure~\ref{fig:sur-vs-mur} (top-row) illustrates the advantage of using \ac{MUR} (with 5 representations) over \ac{SUR} to model the user interests. Specifically, \ac{SUR}-based model is unable to capture multiple interest regions, whereas \ac{MUR} can learn multiple representations to enable this, allowing for better personalization.

While \ac{MUR} can certainly address the desideratum of modeling capacity required for diverse and volatile interests, the overall imbalance in the user interest distribution can hinder the model's ability to learn representations that reflect different interests of the user. Although there have been recent works on different approaches to learn \acl{MUR}~\cite{Li2019Tmall, Cen2020ControllableMF, Kula2017MixtureoftastesMF}, studying the effect of interest/item imbalance on the ability to capture multiple interests with ~\ac{MUR} is largely unexplored. Without taking into account the imbalance problem, existing methods for \ac{MUR} implicitly assume that the so-called interests of the users are equally represented in the dataset. Unfortunately, real-world settings rarely offer a balanced distribution of items/interests. In fact, training models over highly skewed long-tail distributions is one of the most notorious problems prevalent when building recommender systems~\cite{liu2020longtailrecommendation}. Figure~\ref{fig:sur-vs-mur} (bottom-row) highlights the issues observed with imbalanced data. Both \ac{SUR} and \ac{MUR} dominate on the head interests, while completely ignoring other interests. Therefore, an \ac{MUR}-based recommender system can be vulnerable to the imbalanced distribution in the training data. We show that a highly skewed distribution of items/interests causes the gains from using ~\ac{MUR} over ~\ac{SUR} to diminish since the model is unable to capture multiple interests with ~\ac{MUR}.

In this paper, we identify and address the vulnerability of \ac{MUR} to interest imbalance by making a number of technical innovations. We begin by analyzing \ac{MUR} on synthetic datasets generated for the sequential recommendation task. The generative mechanism, despite its simplicity, can be used to control several characteristics that are inspired from the real-world datasets, including sparsity in items and user interests, user interest volatility, and interest exploration. The main motivation behind using a synthetic dataset is to formulate a framework to analyze the effect of interest and item imbalance on the retrieval performance. Using synthetic experiments, we derive two important insights. First, we empirically show that the relative improvement observed by using \ac{MUR} over \ac{SUR} decreases, especially for tails items, as we increase the skewness in the interest distribution. Second, we show, both quantitatively and qualitatively, that a model trained using \ac{MUR} leads to a better clustering structure for item representations compared to a model trained with a \ac{SUR}. However, as we increase the skewness in the data, the quality of clustering structure decreases due to underrepresented tail items in the training dataset.

Our findings from the analysis on synthetic data motivate our key contribution to address the vulnerability of \ac{MUR} to skewness and improve its robustness to data imbalance. We propose a novel density-based weighting approach, referred to as \acf{IDW}, designed to reduce the sensitivity of \ac{MUR} to imbalanced training distribution. \ac{IDW} leverages the item representation space to mitigate the effect of training the recommender model on a highly skewed item/interest distribution. 
In our synthetic experiments, we show that \ac{IDW} improves the robustness of \ac{MUR} to distribution skewness by consistently outperforming the baseline ~\ac{MUR} for overall and tail performance. Moreover, \ac{IDW} leads to an improved clustering structure of the item representations even when the model is trained on a highly skewed interest distribution. On applying the proposed weighting method on real-world datasets, we observe improved performance relative to several other baselines. 

\noindent To summarize, the main contributions of this work are as follows:
\begin{itemize}[leftmargin=1em]
\item A comprehensive analysis on synthetic data, describing the sensitivity of model performance and clustering structure to data skewness for \ac{SUR} and \ac{MUR}. Our analysis depicts the advantage of \ac{MUR} over \ac{SUR} from the standpoint of the structure of the learned item representations, which could be of independent interest to the community when comparing \ac{MUR} and \ac{SUR} models.
\item We propose a novel density-based weighting mechanism that leverage improved item representation structure to make \ac{MUR} robust to imbalanced interest distributions. Combination of density-based weighting with user model calibration results in improved tail item recommendations along with better clustering structure of the learned item representations.
\item Extensive experimental results on three real-world benchmarks show that \ac{IDW} consistently improves tail item recommendations significantly over baseline \ac{MUR}, while also achieving solid improvements for the overall performance in both hit ratio (HR@K) and ranking (NDCG@K) metric. 
\vspace{-0.21em}
\end{itemize}

\section{Related Work}
\textbf{Multi-Interest User Representations.} Extracting multiple interest user representations for recommendation systems was first proposed in \cite{Weston2013NonlinearLF}, where a user is modeled by a set of latent vectors, and the recommendation score of each item is defined as the maximum dot product of the item representation and each latent vector. This was extended in the Embedding-Mixture Factorization (EM-F) and the Projection-Mixture Factorization (PM-F) \citep{Kula2017MixtureoftastesMF} where the item recommendation score is a weighted combination of recommendation scores obtained using each user interest representation. Recently, there has been work on sequential recommender systems where multiple interest representations are used to model a user's item history. In \cite{Kula2017MixtureoftastesMF}, Mixture LSTM (M-LSTM) was proposed that combines the LSTM network with the PM-F model to extract multiple user representations in a sequential setting. ComiRec~\cite{li2019multiinterest} and MIND~\cite{Cen2020ControllableMF} are recent methods that propose multi-interest extractor layers based on the capsule routing mechanism. 
Although there is a vast amount of work that has focused on model architectures for \ac{MUR}, the robustness of ~\ac{MUR} to imbalanced user interest distributions is largely unexplored. We analyze the effect of skewness in user interests on ~\ac{MUR} and propose a novel density-based weighting scheme for improving the recommendation performance of tail items with \ac{MUR}. While this paper focuses on a transformer encoder-based architecture to model ~\ac{MUR} (details in Section~\ref{sec:preliminaries}), our work nicely complements prior advances made in modeling ~\ac{MUR} since the proposed density-based weighting can be used with other multi-interest frameworks~\cite{li2019multiinterest, Cen2020ControllableMF, Kula2017MixtureoftastesMF} proposed for recommendation systems.

\noindent\textbf{Long-tail Learning.}
In most real-world datasets, the data distribution is highly skewed and often follows a long-tail distribution. Learning models on long-tail distribution has been studied across many areas including computer vision~\cite{cui2019class, Lin2018FocalLoss}, natural language processing~\cite{czarnowska2019don}, and recommendation systems~\cite{liu2020longtailrecommendation}. In recommendation datasets, the long-tail distribution is especially apparent due to the large scale of items and highly skewed user interactions. The recommender systems are heavily influenced~\cite{Domingues2012Music, Beutel2017Beyond} by the long-tail distribution, leading to poor performance for the tail items.

There are several approaches that can allow learning from imbalanced data. Recent methods~\cite{wang2017learntail, Liu2019Openlong, Cui2018iNatTransfer} that perform head-to-tail knowledge transfer have gained attention in computer vision. Knowledge transfer has also been extended in recommendation systems~\cite{zhang2021model}. Approaches to tackle cold-start recommendation~\cite{zhu2018addressing, liang2020joint, Dong2020Mamo, lee2019melu} can also improve tail item prediction. Aforementioned approaches rely on additional side information (item and user content features), which can be difficult to acquire, especially when there are heterogeneous items~\cite{zhu2021crossdomain} that do not share a common feature space. In this work, we only focus on approaches that do not depend on additional side information.

Frequency-based regularizers~\cite{yi2019samplingbias, menon2021longtail, Bengio2008AdaptiveIS} have been used to reduce the sampling bias when training models on long-tail distribution. Some methods modify the loss objective~\cite{Lin2018FocalLoss, cui2019class, Beutel2017Beyond, he2009Imbalance} using sample weighting for each class/item in the dataset. However, these methods use the item frequencies in the training dataset to determine the weights, which can lead to sub-optimal performance for sequential recommendation systems as item frequencies alone lack the contextual information contained in user's past history. Unlike the above methods, our proposed density weighting uses the learned item representations that encapsulate the contextual information from the user's history to determine and mitigate the effective imbalance. Closest to our method is the recent work~\cite{Steininger2021DensityWeighting, Yang2021ImbalancedRegression} used in regression models, which uses density estimation in the target space to mitigate imbalance. However, in contrast to the regression datasets that have fixed labels, in recommendation system, the labels for the user tower ($i.e.$ item representations) are also learned using a separate model, making it a more challenging task. We make several contributions to enable density weighting for recommendations.

\section{Methods}
\label{sec:methods}
\textbf{Problem Statement.} Consider a set of users $\gU$ and items $\gI$, where $u \in \gU$ denotes a user and $i \in \gI$ denotes an item. We will use the notation $|\gU|$ and $|\gI|$ to denote the total number of users and items respectively. Given the user's historical behaviour in the form of a sequence of item engagements $x_u = \{i^{(u)}_1, \dots, i^{(u)}_n\}$, our recommendation task is to predict the next item that the user is likely to interact with at the $(n+1)^{th}$ step. For ease of notation, we will denote the training dataset as $\gD = \{x_u, y_u\}_{u=1}^{|\gU|}$, where $y_u$ refers to the $i^{(u)}_{n+1}$ item and is considered the label for the user $u$.

\subsection{Background}
\label{sec:preliminaries}
\begin{figure}[!t]
    \includegraphics[width=0.4\textwidth]{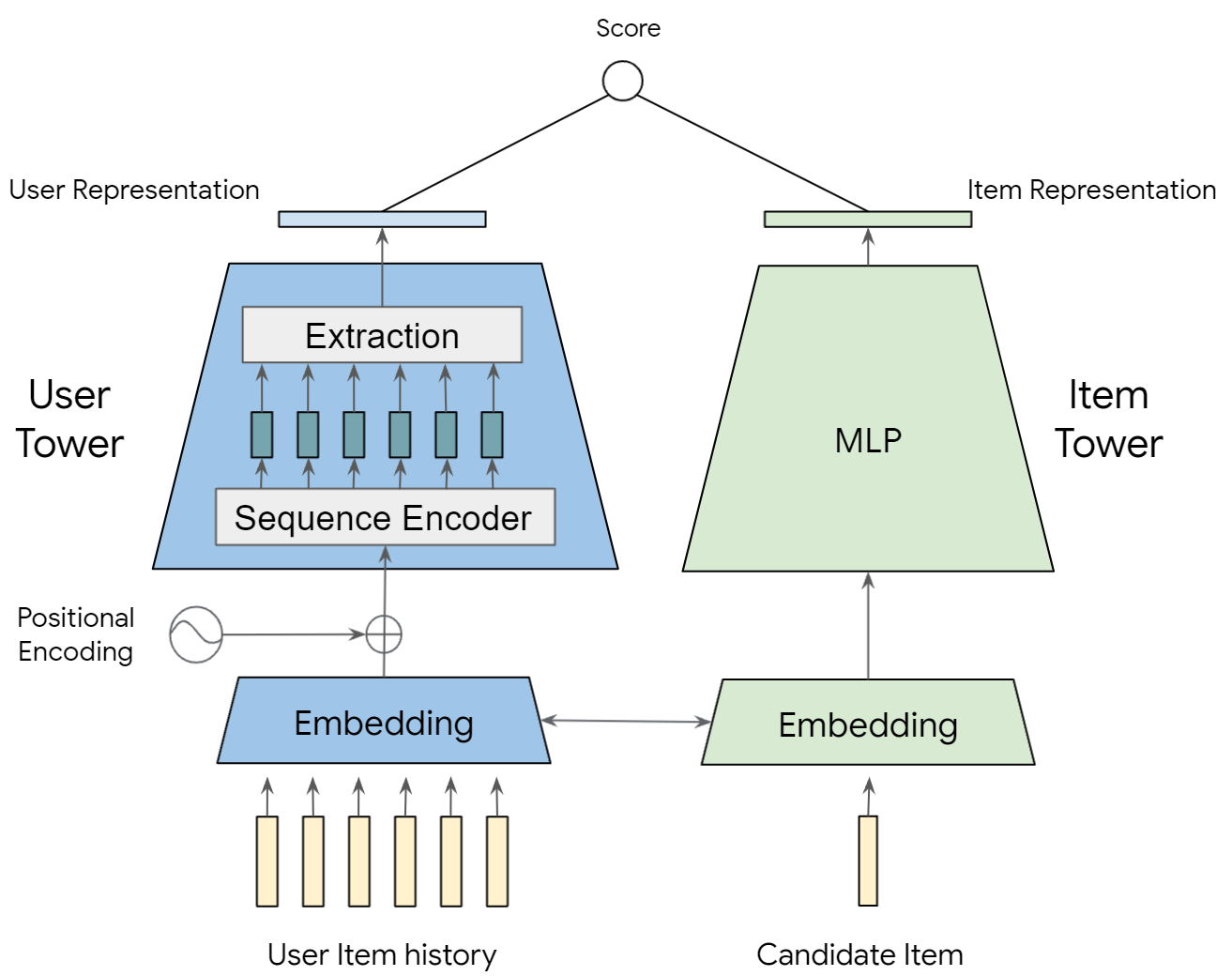}
    \caption{A typical two-tower DNN model for next item prediction. (Left) User tower that outputs a single user representation. (Right) Item tower.}
    \label{fig:two_tower}
    \vspace{-2em}
\end{figure}

\textbf{Neural Deep Retrieval Model.}
Motivated by the success of deep learning in computer vision and natural language processing, applying deep neural networks (DNNs) for recommender systems is prevalent. Deep representation learning provides a powerful framework to encode the users and the items in a low-dimensional representation space, which can be used for retrieval in recommender systems. We apply the two-tower DNN model that aims to learn a common low-dimensional representation space for users and items. Figure~\ref{fig:two_tower} provides an illustration of a typical two-tower model architecture used for next item prediction, where the left and right tower encode the user item history and the candidate items respectively. A user tower consists of a sequence encoder ($e.g.$ Transformer, LSTM, $etc$.) and a representation extraction layer that reduces the outputs of the sequence encoder to a single user representation ($e.g.$ by weighted aggregation or by picking the hidden output of the encoder from the latest time step). The item tower is a multi-layer perceptron (MLP) model that learns the representation of item candidates given a one-hot encoding of the item. The towers can also easily incorporate side-information of users and items, if available. We denote the user tower as $f_\theta$ and the item tower as $g_\phi$, where $\theta$ and $\phi$ are the set of trainable parameters of the two respective architectures. The two-tower model uses the batch softmax optimization~\cite{Bengio2008AdaptiveIS, Covington2016Youtube} for training. The item probabilities $P\left(y|x_u; \theta, \phi \right)$ are calculated over a random batch:
\begin{small}
\begin{align}
    P\left(y|x_u; \theta, \phi \right) &= \frac{exp\left(s\left(f_\theta(x_u), g_\phi(y)\right)\right)}{\sum_{y' \in \gB} exp\left(s\left(f_\theta(x_u), g_\phi(y')\right)\right)}
\end{align}
\end{small}
where $\gB$ is the set of item labels in the sampled batch and $s(a, b)$ is the affinity score defined as the standard dot product. Naively using the batch softmax probabilities leads to a biased-estimate of $P\left(y|x_u; \theta, \phi \right)$, where popular items are overly penalized as negatives due to high probability of being included in the batch. To mitigate this bias, LogQ correction~\cite{jean2014using, yi2019samplingbias} is used with the sampled softmax loss, where the logit scores are corrected as follows: 
\begin{small}
\begin{align}
s^c\left(f_\theta(x_u), g_\phi(y)\right) = s\left(f_\theta(x_u), g_\phi(y)\right) - \log (p(y)).    
\end{align}
\end{small}
Here p(y) denotes the sampling probability of item $y$ in a random batch. Given the softmax probabilities, the recommendation task can be formalized as a multi-class classification problem with $y_u$ as the label for the input $x_u$, and the negative log-likelihood ($\gL(\theta, \phi)$) of the training dataset can be defined as:
\begin{small}
\begin{align}
\gL(\theta, \phi) = \frac{1}{|\gU|}\sum_{\{x_u, y_u\} \in \gD} - \log{P\left( y_u | x_u; \theta, \phi \right)}.
\label{eq:log-likelihood}
\end{align}
\end{small}
The models are trained to minimize $\gL(\theta, \phi)$ with respect to the parameters $\theta$ and $\phi$.  Thus, the affinity score for the user representation and the corresponding label item (also referred to as the ``positive item'') is increased, while the affinity for the user representation and other item labels in the mini-batch (referred to as the ``negative items'') is reduced. The optimization procedure implicitly pulls items that correspond to the same user interest close to each other in the representation space, while items belonging to different interests are pushed away.\\
\begin{figure*}[!h]
    \captionsetup[sub]{font=small,labelfont={bf,sf}}
    \centering
    \begin{subfigure}[b]{0.16\textwidth}
      \includegraphics[width=\linewidth,height=\linewidth]{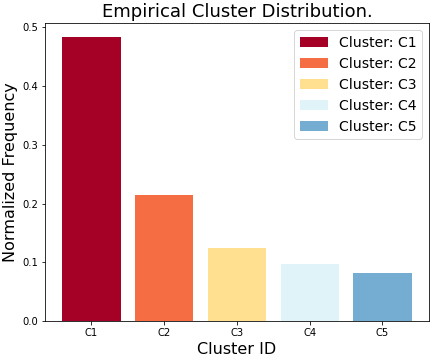} 
      \caption{Sampled interests.}
     \label{fig:syn-cluster-distrib}
    \end{subfigure}
    \begin{subfigure}[b]{0.16\textwidth}
      \includegraphics[width=\linewidth,height=\linewidth]{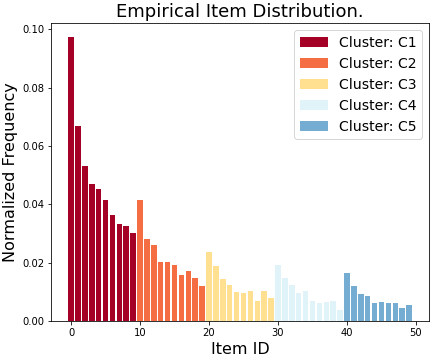}
      \caption{Sampled items.}
     \label{fig:syn-cond-item-distrib}
    \end{subfigure}
    \begin{subfigure}[b]{0.16\textwidth}
      \includegraphics[width=\linewidth,height=\linewidth]{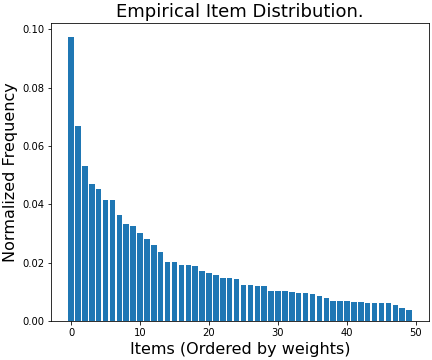}  
      \caption{Item distribution.}
      \label{fig:syn-item-distrib}
    \end{subfigure}
    \begin{subfigure}[b]{0.24\textwidth}
      \includegraphics[width=\linewidth]{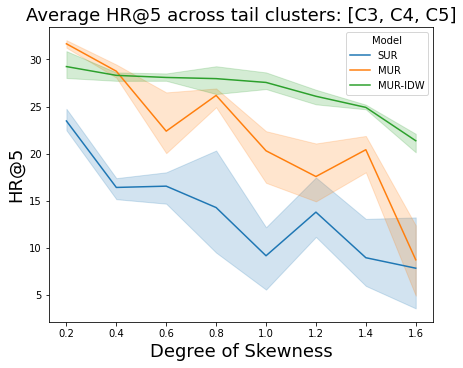}  
      \caption{\% HR@5 for [C3, C4, C5].}
      \label{fig:syn-tail-clusters}
    \end{subfigure}
    \begin{subfigure}[b]{0.24\textwidth}
      \includegraphics[width=\linewidth]{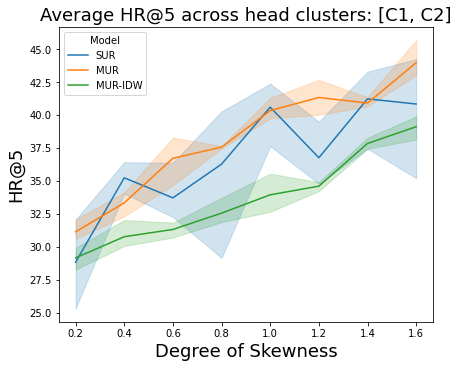}  
      \caption{\% HR@5 for [C1, C2].}
      \label{fig:syn-head-clusters}
    \end{subfigure}\\
    \vspace{0.5em}
    \begin{subfigure}{0.24\textwidth}
      \includegraphics[width=\linewidth]{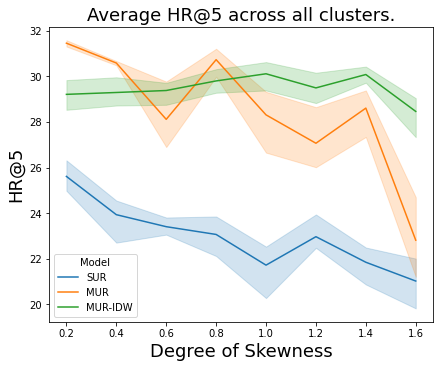} 
      \caption{\% HR@5 for all clusters.}
      \label{fig:syn-avg-clusters}
    \end{subfigure}
    \begin{subfigure}{0.24\textwidth}
      \includegraphics[width=\linewidth]{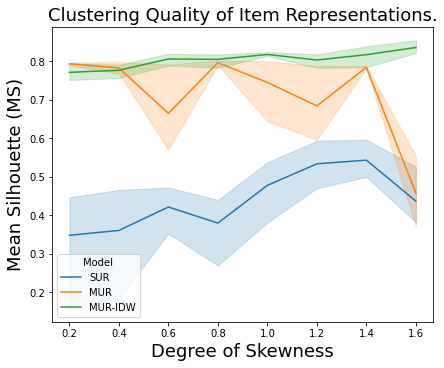}
      \caption{Clustering Quality.}
      \label{fig:syn-ms-plot}
    \end{subfigure}
    \begin{subfigure}{0.24\textwidth}
      \includegraphics[width=\linewidth]{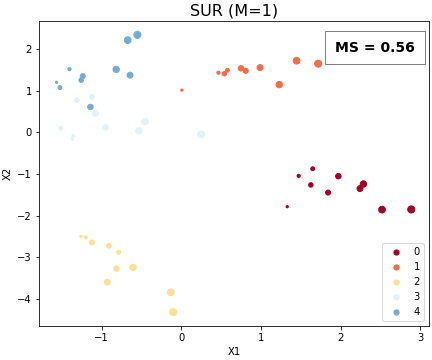}  
      \caption{SUR Synthetic}
      \label{fig:sur-synthetic}
    \end{subfigure}
    \begin{subfigure}{0.24\textwidth}
      \includegraphics[width=\linewidth]{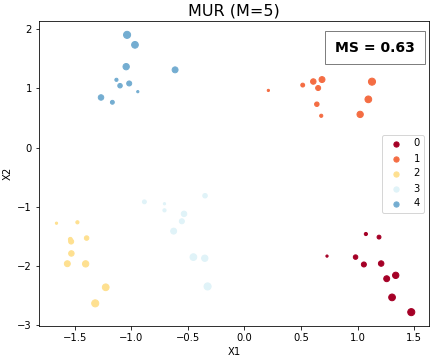}  
      \caption{MUR Synthetic}
      \label{fig:mur-synthetic}
    \end{subfigure}
    \vspace{-1em}
    \caption{Analysis on the synthetic data having 50 items, 5 user interests (clusters) and varying degree of skewness in the interest distribution.}
    \label{fig:syn-results}
    \vspace{-1em}
\end{figure*}

\noindent\textbf{Module for Multi-Interest Extraction.} We extend the standard user tower described in Figure~\ref{fig:two_tower} to model \ac{MUR} using a multi-interest extraction layer. The goal of such a layer is to output \acl{MUR}, $i.e.$ $f_\theta(x_u) = \left\{z_u^{(1)}, \dots, z_u^{(M)}\right\}$, where $M$ denotes the number of user representations extracted. While $\{z_u^{(m)}\}_{m=1}^M$ could be learned separately using $M$ separate user towers, such an approach will be inefficient in both computation and the number of parameters. \citet{Humeau2020PolyEncoders} recently proposed an efficient and scalable approach that uses global attention to extract multiple context embeddings for the natural language sentence matching tasks. We adapt this idea of global attention for a recommender system to extract multiple representations for a given user history. Let the output of the sequence encoder be $\{h^{(1)}_u, \dots, h^{(n)}_u\}$, then to obtain multi-interest user representations, we learn $m$ global query vectors $\{q_1, \dots, q_M\}$, where each $q_m$ extracts the user representation $z_m$ using attention:
\begin{small}
\begin{align}
    z_u^{(m)} = \sum_{i=1}^n \alpha^{(i)}_{m} h^{(i)}_u \quad \text{where} \quad \alpha^{(i)}_{m} = \frac{exp\left(q_m^T h_u^{(i)}\right)}{\sum_{i=1}^n exp\left(q_m^T h_u^{(i)}\right)}
    \nonumber
\end{align}
\end{small}
Given multiple user representations, we compute the affinity score for each candidate $y$ as follows: 
\begin{small}
\begin{align}
    s\left(f_\theta(x_u), g_\phi(y)\right) &= {\left({z_u}^T g_\phi(y)\right)}, \enspace \text{where}\\
    z_u = \sum_m w^{(m)} {z^{(m)}_u},  \quad &
    w^{(m)} = \frac{exp\left({ z^{(m)}_u}^T g_\phi(y) \right)}{\sum_{j=1}^m exp\left( { z^{(j)}_u}^T g_\phi(y) \right)} \nonumber
\end{align}
\end{small}

As the above multi-interest extraction approach uses global query parameters in attention, we refer this representation extraction layer as parametric attention. Note that there are alternate methods for multi-interest extraction layers~\cite{Cen2020ControllableMF, Kula2017MixtureoftastesMF, Li2019Tmall}, but in this work we only consider parametric attention to extract \ac{MUR} due to its simplicity and scalability. More importantly, the focus of this work is on studying the effect of long-tail distribution on \ac{MUR} in the two-tower DNN architecture~\cite{Covington2016Youtube}, which has become a popular framework for recommendations in the recent years. 

\subsection{Synthetic Data Analysis}
\label{section:synthetic-analysis}
In this section, we discuss our analysis on the synthetic data, which forms the basis of our density weighting scheme. The synthetic data for sequential recommendation is generated using a finite state hidden Markov model~\cite{baum1966markov, Juang1991HMMSpeech}, where we consider the hidden state clusters as the so-called user interests. To define the Markovian process for the sequence, we assume a user-specific interest transition matrix which is a function of the user interests, distribution skewness, and user interest volatility i.e., probability of switching between different user interests. A detailed description of the data generating process is provided in Appendix~\ref{apx:synthetic_data}. Figure ~\ref{fig:syn-cluster-distrib}-\ref{fig:syn-item-distrib} show examples of synthetically generated data distributions. The synthetic data models long-tailed distributions for both interest and item space to capture the power-law behavior of items encountered in real recommendation systems. 

While the Markovian assumption may not reflect complex real-world scenarios, we find that our observations are consistent in real-world datasets as described later in Section~\ref{sec:experiments}. Our key motivation to use synthetic data is to study the effect of distribution skewness on modeling capabilities of \ac{MUR}. Moreover, item clusters in the synthetic data can be treated as ground-truth interests for the users, allowing us to quantitatively (using ~\ac{MS} score~\cite{Rousseeuw1987Silhouette}) and qualitatively (See Figure~\ref{fig:sur-vs-mur}) evaluate the structure of the learned item representations for different models.

\noindent\textbf{Retrieval performance.} To study the sensitivity with respect to an imbalanced distribution, we evaluate the performance of the models trained with varying degree of data skewness. Specifically, we vary the imbalance in the user interest distribution and plot the recommendation performance (Hit Ratio, expressed as \%) across different splits: tail/infrequent clusters (Figure~\ref{fig:syn-tail-clusters}), head/frequent clusters (Figure~\ref{fig:syn-head-clusters}), and all clusters (Figure~\ref{fig:syn-avg-clusters}). As seen in Figure~\ref{fig:syn-tail-clusters}, increasing the skewness causes the performance of \ac{MUR}-based model to decrease for tail interests, which also results in decreasing average performance across all the interest clusters (Figure~\ref{fig:syn-avg-clusters}). In fact, when the user interests are highly skewed, the performance of \ac{MUR}-based model becomes comparable to the performance obtained using the \ac{SUR}-based model. This is due to the under-representation of items in the tail interest clusters, causing the model to dominate on the head interest items. On the other hand, using our proposed density weighting with \ac{MUR}, referred to as \ac{MUR}-IDW (described in Section~\ref{sec:idw}), makes the model robust to imbalance as shown in Figure~\ref{fig:syn-tail-clusters} and Figure~\ref{fig:syn-avg-clusters}.\\

\par\noindent\textbf{Clustering quality.} We inspect the structure of the representation space learned using different models. A good representation space would have localized clusters representing user interests, such that the items belonging to a similar interest have item representations close to each other. First, we do a qualitative analysis of the clustering structure learned using \ac{MUR} and \ac{SUR}. Figure~\ref{fig:sur-synthetic}-~\ref{fig:mur-synthetic} shows that \ac{MUR} leads to a better clustering structure of the representation space compared to the representations learned with \ac{SUR} on the synthetic data.

Next, we measure the clustering quality of the item representations using the \acf{MS}~\cite{Rousseeuw1987Silhouette} score. The \ac{MS} score measures the average tightness and separation of the clusters. For a point $p$ belonging to the ground truth cluster $c$, we define the silhouette as $s_p = \frac{b_p - a_p}{max(a_p, b_p)}$, where $a_p$ is the mean distance between $p$ and other points in the cluster $c$, and $b_p$ is the mean distance between $p$ and the points in the closest cluster. Specifically, we can calculate $a_p$ and $b_p$ as:
\begin{small}
\begin{align}
    a_p = \frac{1}{|c|-1} \sum_{q\in c, p \neq q} dist(p, q);
    \quad b_p = \min_{\tilde{c} \neq c}  \frac{1}{|\tilde{c}|} \sum_{q \in \tilde{c}} dist(p,q) \nonumber
\end{align}
\end{small}
The MS score is computed by taking the average over all points, $i.e$ $MS := \E_p\left[s_p\right]$. The \ac{MS} score takes the value $\in[-1, 1]$, where a higher MS depicts distinct and separated clusters, and a lower MS depicts overlapping clusters. Figure~\ref{fig:syn-ms-plot} shows the \ac{MS} score for varying degrees of skewness in the synthetic dataset. While the clustering structure obtained with \ac{MUR} is consistently better than with \ac{SUR}
, the clustering quality in the representation space deteriorates with increasing degree of imbalance.\\ 

\vspace{-1em}
\subsection{Iterative Density Weighting}
\label{sec:idw}

In this section, we describe our novel density-weighted loss for the two-tower recommendation system, which advocates \textit{leveraging the structure of the item representation space} to learn and mitigate the effective imbalance of the items. This proposal aims to leverage the item representation space enabled by \ac{MUR} to enhance the overall recommendation performance.

To determine the effective item imbalance, we use item distribution smoothing that convolves a symmetric kernel with the empirical distribution of the items. This results in a kernel-smoothed version of the item distribution that accounts for the overlap in information of nearby items in the representation space. We then use this kernel-smoothed item distribution to learn a balanced loss that is uniformly weighted in the item representation space.

We first consider the simple frequency-balanced~\cite{he2009Imbalance} version of the loss function described in (\ref{eq:log-likelihood}):
\begin{small}
\begin{align}
\gL_{BAL}(\theta, \phi) = \frac{-1}{|\gU|} \sum_{\{x_u, y_u\} \in \gD} \frac{1}{p(y)} \log{P\left( y_u | x_u; \theta, \phi \right)}, \text{ where}
\label{eq:log-likelihood-item-balanced}
\end{align}
\end{small}
$p(y) := n_y / |\gU|$ corresponds to the empirical distribution of the items and $n_y$ denotes the item frequency. Minimizing the balanced loss in ~(\ref{eq:log-likelihood-item-balanced}) aims to learn a uniform loss function across all the items in the training dataset. However, the imbalance in $p(y)$ may not reflect the effective imbalance of the items in the representation space, and naively optimizing ~(\ref{eq:log-likelihood-item-balanced}) can lead to overfitting of the model to tail items that have low $p(y)$. As we show in Section~\ref{sec:experiments}, such a frequency-based balanced loss can lead to poor overall recommendation performance. 

In contrast, our density-weighted loss uses the kernel-smoothed version of the empirical item distribution:
\begin{small}
\begin{align}
\tilde{p}(y) := \int k(y, y') \, {p(y')} dy', \text{ where}
\end{align}
\end{small}
$k(y, y')$ is the symmetric Gaussian kernel that characterizes the distance of items in the representation space. Therefore, $\tilde{p}(y)$ leverages the structure of the representation space in determining the effective imbalance of items. In particular, $\tilde{p}(y)$ captures the effective weight of the region defined by the neighborhood of item $y$ in the representation space, thereby reducing the model's sensitivity to $p(y)$ through kernel smoothing. Note that since $\tilde{p}(y)$ is determined using the item representations learned in the item tower, in practice we use $\tilde{p}(y|\phi)$. As both the kernel-smoothed density $\tilde{p}(y|\phi)$ and the loss used to train the parameters of the two-tower recommender model $\{\theta, \phi$\} are inter-dependent, we propose an iterative two-phase procedure (Algorithm~\ref{alg:idw}) to train the two-tower architecture. Given the model parameters $\{\theta^{(t)}, \phi^{(t)}\}$ and the density $\tilde{p}(y|\phi^{(t)})$ at iteration $t$,
\begin{itemize}[leftmargin=*]
\setlength{\partopsep}{4pt}
\item In the Sleep Phase, the item representations from the item tower are used to update the density through kernel-smoothing. 
\item In the Wake Phase, the two-tower model is trained with the balanced loss given by the item density from the previous iteration.
\end{itemize}

\vspace{1em}
\noindent\textbf{Sleep Phase: Update Item Weights.} To learn a balanced loss function with respect to $\tilde{p}(y|\phi^{(t)})$, we introduce item-specific weights $w_y^{(t)}$, such that the initial $w_y^{(0)} = n_y/|\gU|$. The sleep phase updates the item weights based on the item representations defined by the current state of the item tower. In particular, the density update is defined as:
\begin{small}
\begin{align}
    \tilde{p}(y| \phi^{(t)}) &= \frac{1}{\ell} \sum_{k=1}^{|\gI|} w_k^{(t)} \gK\left( \frac{||y - v_k||}{\ell} \right), \text{ where}
    \label{eq:kde}
\end{align}
\end{small}
$v_k = g_{\phi^{(t)}}(i_k)$ refers to the representation of the $k^{th}$ item, $\gK$ is the Gaussian kernel function with the bandwidth parameter $\ell$ (the bandwidth $\ell$ is selected based on Scott's Rule~\cite{Scott2012MultivariateDE}, which is a standard bandwidth selection method). In practice, we found using relative normalized density~\cite{Steininger2021DensityWeighting} as a better estimate:
\begin{small}
\begin{align}
    p'(y|\phi^{(t)}) &= \frac{\tilde{p}(y| \phi^{(t)}) - \min_{y}\tilde{p}(y| \phi^{(t)})}{\max_y \tilde{p}(y| \phi^{(t)}) - \min_{y}\tilde{p}(y| \phi^{(t)})}
    \label{eq:relative-density}
\end{align}
\end{small}
The relative item density of each item is $\in$ $[0,1]$ and the item weights are updated with $h^{(t)}_k := 1 - p'(y=k|\phi^{(t)})$ resulting in a negative correlation with the relative item density. To avoid a drastic change in the loss function, the weights are updated gradually using a momentum parameter $m$:
\begin{small}
\begin{align}
    w_k^{(t+1)} = m \cdot w_k^{(t)} + (1-m) \cdot  {\left(h^{(t)}_k / \sum_{{k'}} h^{(t)}_{k'} \right)}\label{eq:update_weights}
\end{align}
\end{small}
where we use the subscript $k$ to denote the $k^{th}$ item. Since the updates are negatively correlated with the item density, they allow the model to progressively upweight/downweight the regions with low/high item density respectively. The complete procedure followed in the sleep phase is summarized in Algorithm~\ref{alg:compute_weights}.
\\

\noindent\textbf{Wake Phase: Train Two-Tower Model.} Given $w^{(t+1)}_{y}$, the two-tower model is trained by minimizing the following weighted loss:
\begin{small}
\begin{align}
    \tilde{\gL}_{BAL}(\theta, \phi) &= \frac{-1}{|\gU|} \sum_{\{x, y\} \in \gD}  w^{(t+1)}_{y} \cdot \log{P \left( y | x; \theta^{(t)}, \phi^{(t)} \right)}.\\
    \{\theta^{(t+1)}, \phi^{(t+1)}\} &= \argmin_{\theta, \phi} \tilde{\gL}_{BAL}(\theta, \phi)
    \label{eq:weighted-log-likelihood}
\end{align}
\end{small}

{
\setlength{\textfloatsep}{0pt}
\begin{algorithm}[!t]
\caption{\small Iterative Density Weighting}\label{alg:idw}
\begin{algorithmic}[1]
    \State {\bf Input:} Training Dataset $\gD$, Weight momentum $m$.
    \State \textbf{Initialize}: $w_k^{(0)} \gets n_k/|\gU|$ \Comment{Normalized Item Frequency}.
    \State \textbf{Initialize}: $\theta^{(0)}, \Phi^{(0)}$ \Comment{Trained with unweighted loss~(\ref{eq:log-likelihood})}.
    \For  {$t \in 0, \dots, MAX_T$}
        \State \textbf{Sleep:} $w^{(t+1)} \gets \text{UpdateWeights}(m, w^{(t)}, \phi^{(t)})$
        \If{$||w^{(t+1)} - w^{(t)}|| < \eta$}
            \State break
        \EndIf
        \State \textbf{Wake:} $\{\theta^{(t+1)}, \phi^{(t+1)}\} \gets$ TrainTowers($w^{(t+1)}$).
    \EndFor
    \State $\phi^{*} \gets \phi^{(t)}$
    \State $\theta^{*} \gets$ UserTowerCalibration($\theta^{(t)}, \phi^{*}$)
\end{algorithmic}
\end{algorithm}

\begin{algorithm}[!t]
\caption{\small Sleep Phase: Update Item Weights.}\label{alg:compute_weights}
    \begin{algorithmic}[1]
\Function{UpdateWeights}{$m$, $w^{(t)}$, $\phi^{(t)}$}
    \State $v_k \gets$ $g_{\phi^{(t)}}\left(i_k\right)$. \Comment{Item tower forward pass.}
    \State Obtain $p(y| \phi^{(t)})$ using Equation~(\ref{eq:kde}).
    \State Obtain $p'(y|\phi^{(t)})$ using Equation~(\ref{eq:relative-density}).
    \State $h^{(t)}_k \gets 1 - p'(y = k|\phi^{(t)})$.
    \State Obtain $w_k^{(t+1)}$ using equation~(\ref{eq:update_weights}).
    \State \Return $w^{(t+1)}$
\EndFunction
\end{algorithmic}
\end{algorithm}
}
\begin{figure*}[!h]
    \centering
    \begin{subfigure}{0.8\textwidth}
      \includegraphics[width=\linewidth]{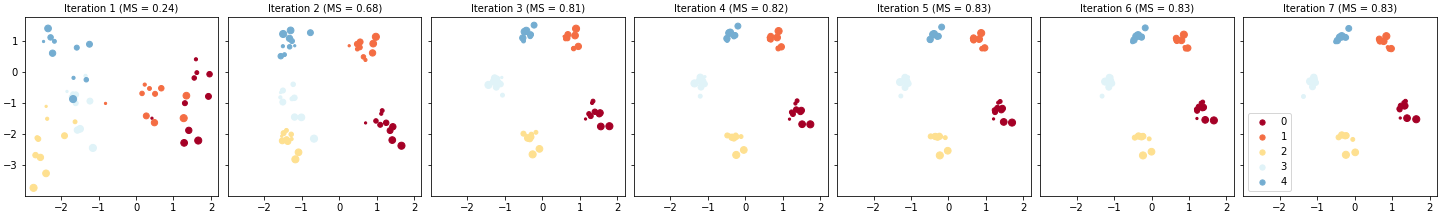}
    \end{subfigure}\\
    \begin{subfigure}{0.8\textwidth}
      \includegraphics[width=\linewidth]{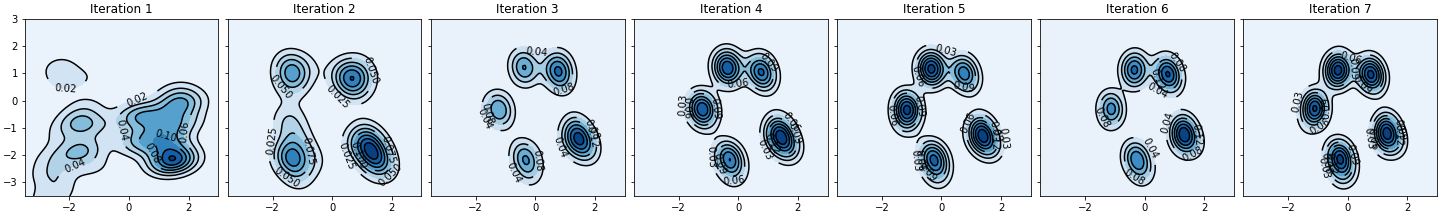}
    \end{subfigure}
    \vspace{-1em}
    \caption{The evolution of item embedding space and item distribution over the representation space over several iterations of \ac{IDW}. The improvement in the clustering structure of the item embeddings is illustrated in the first row and also captured by the MS score, increasing from $0.24$ at iteration 1 to $0.83$ in iteration 7.}
    \label{fig:IDW_iterations}
    
\end{figure*}

\noindent\textbf{Stopping Criterion.} To detect convergence in \ac{IDW}, we track $||w^{(t+1)} - w^{(t)}||$. As the model is trained, the loss contributions becomes uniform in the item representation space. We visualize the evolution of the item representation space for the synthetic data in Figure~\ref{fig:IDW_iterations}. We observed that making the loss uniform in the representation space, progressively improves the clustering structure of the item representations. The improvement in the clustering structure is also reflected in the \acf{MS} score in Figure~\ref{fig:IDW_iterations} (Top).\\

\noindent\textbf{User Tower Calibration.} \label{section:user-tower-calibration} 
While the uniform loss structure improves the clustering structure of the items, the retrieval performance over the head clusters can drop compared to when the model is trained using the unweighted likelihood. Therefore, to mitigate the performance drop for head clusters, once we improve the item clustering structure using \ac{IDW}, we freeze the item tower and train only the user tower with the unweighted loss. We study the importance of user tower calibration in section~\ref{sec:ablation-idw-calibration}.

\section{Experiments}
\label{sec:experiments}
\input{Tables/imbalance_overall_k20}
\begin{figure*}[!ht]
    \centering
    \begin{subfigure}[b]{0.22\linewidth}
      \centering
      \includegraphics[width=\linewidth]{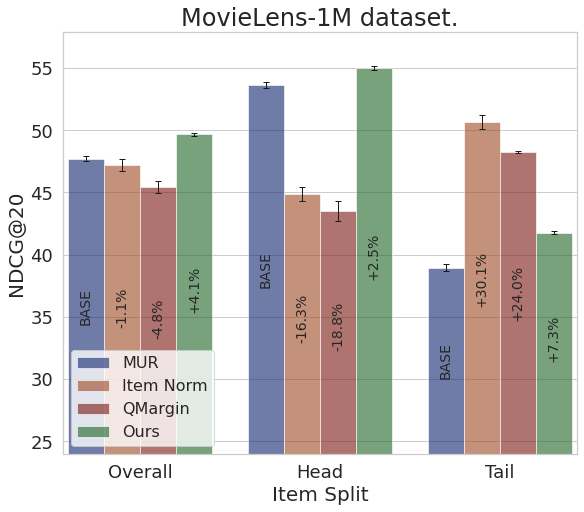} 
     \label{fig:ndcg20_movielens_split}
    \end{subfigure}
    \qquad
    \begin{subfigure}[b]{0.22\linewidth}
      \centering
      \includegraphics[width=\linewidth]{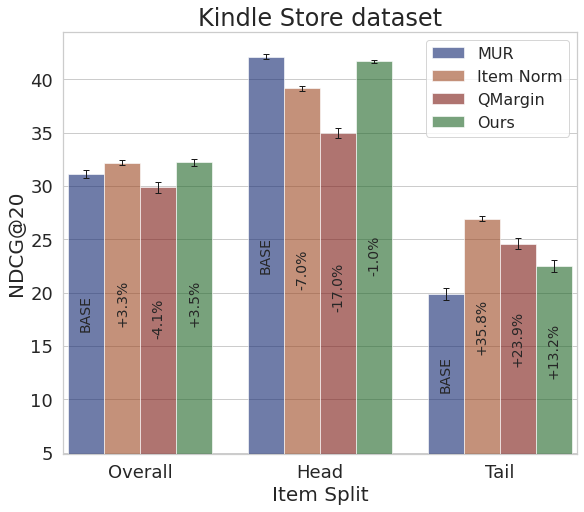} 
     \label{fig:ndcg20_kindle_split}
    \end{subfigure}
    \qquad
    \begin{subfigure}[b]{0.22\linewidth}
      \centering
      \includegraphics[width=\linewidth]{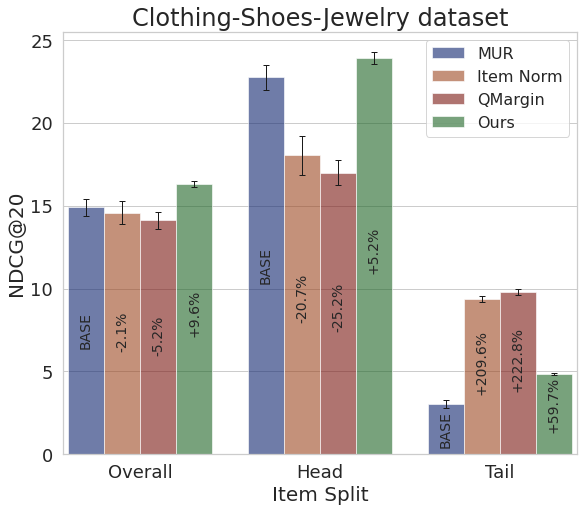} 
     \label{fig:ndcg20_clothing_split}
    \end{subfigure}
    \vspace{-2em}
    \caption{NDCG@20 on the overall, head and tail item sets for three real-world benchmarks. The relative increase/decrease in the performance is calculated relative to the MUR (BASE) performance on the respective item set.}
    \label{fig:head_tail_split_results}
    \vspace{-1em}
\end{figure*}
\subsection{Datasets}
We evaluate the proposed method on three public benchmarks that are used for recommender models. The statistics of the datasets are summarized in Table~\ref{table:dataset_summary}. In all the three datasets, the item distribution follows a highly skewed long-tail distribution (See the item distributions in Appendix~\ref{apx:item_distr}.). For each benchmark dataset, we preprocess the data to construct item sequences using the history of items that users have interacted with. We consider a maximum sequence length of 30, where padding is added if the item sequences are smaller than 30, and the item sequences larger than 30 are truncated and treated as multiple sequences. 
\input{Tables/dataset_summary}
\subsection{Evaluation Metrics} We employ top-k Hit Ratio (HR@k) and NDCG@K with $k=20$ to evaluate the recommendation performance of all the models. Additional results with $k=5$ are included in the Appendix~\ref{apx:results_k5}.
As the goal of \ac{IDW} is to improve the performance of the tail items without hurting the overall performance across all the items, we further compare the metrics on the tail and head item sets separately. 
Assuming the Pareto Principle~\cite{Box1986Factorials}, the 20\% most frequent items are considered head items, and the remaining items as the tail items. Following the standard evaluation protocol~\cite{kang2018SelfAttentive, Ren2020Sequential}, we use the leave-one-out strategy for evaluation. For each item sequence, the last item is used for testing, the item before the last is used for validation, and the rest is used for training. As the item set is typically large, we use randomly sampled negative items~\cite{kang2018SelfAttentive,
He2017NeuralCollab, koren2008Factorization, Ren2020Sequential} at evaluation. Specifically, we pair the ground-truth item with 99 randomly sampled negative items that the user has not previous interacted with to compute the evaluation metrics. 

\subsection{Baselines}
To demonstrate the effectiveness of the proposed \ac{IDW} on long-tail item distribution, we compare \ac{IDW} with other commonly used weighting methods that deal with imbalance in recommender models.
All the baselines use the two-tower model as the backbone, which has proven to be scalable because of efficient inference through inner products. For extracting multiple user representations to capture diverse user interests, we use parametric attention as described in Section~\ref{sec:preliminaries}. We list all the baselines below. We note that methods that rely on knowledge transfer of user/item features~\cite{Beutel2017Head2Tail, zhang2021model} or augmentation methods (over/under sampling~\cite{brownlee2020imbalanced}) to mitigate imbalance can be easily combined with all the following baselines. Our proposed density-weighting and the baselines considered do not depend on user/item features. 
\begin{itemize}[leftmargin=1em]
\setlength{\parskip}{0pt}
\item \textbf{SUR}: We use a single user representation as the output of the user tower by using $M=1$ in the parametric attention~\cite{Humeau2020PolyEncoders} layer. 
\item \textbf{MUR}: The multiple user representations are extracted using the parametric attention~\cite{Humeau2020PolyEncoders} layer with $M=5$ output representations. We do an ablation to study the effect of $M$ in Section~\ref{sec:ablation-M}.
\item \textbf{Frequency Balance}: We use the frequency-based balancing~\cite{he2009Imbalance} as described in Equation~\ref{eq:log-likelihood-item-balanced}. Note that this is a static weighting scheme, where the item weights do not change during training.
\item\textbf{Focal Balance}: A recent work by~\citet{Lin2018FocalLoss} modifies the loss function to train the model on an imbalanced dataset by down-weighting the loss of well-classified examples. This is a dynamic weighting scheme, where the item weights depend on the loss of each input during training.
\item \textbf{Qmargin}: \citet{CaoMargin2019} proposed  to add a per-class (per-item) margin $\delta_y$ into the cross-entropy loss, where $\delta_y \propto p(y)^{-1/4}$.
\item \textbf{Item Norm}: \cite{kim2020Imbalance} upweights the score of rare items by normalizing the item representations. We also experiment with a post-hoc version of normalized item, where a standard two-tower model is trained and the item representation normalization is only carried out during the inference.~\cite{Kang2020Decoupling}.
\end{itemize}

\subsection{Hyper-parameter Settings} We set the embedding dimension as 16 across all the models for the three benchmarks. We use the two-tower architecture as the backbone for all the models, where the user tower is a self-attention encoder~\cite{Vaswani2017Attention} with two multi-head attention layers and four attention heads per layer. The item tower is an MLP with two fully connected layers each with 16 output nodes. For \ac{IDW}, the momentum update of item weights (Equation~\ref{eq:update_weights}) and $\eta$ are determined using grid search in the range of $\{0.7, 0.8, 0.9, 0.99\}$ and $\{1.0, 2.0, 3.0, 4.0\}*10^{-4}$. We use Adam to optimize all the models with a learning rate of 0.01. For all models, we used early stopping based on the validation dataset. The models are implemented with TensorFlow Recommenders (TFRS)~\cite{tensorflow-recommenders}.
\subsection{Recommendation Performance}
\subsubsection{Overall Performance}
We compare our proposed method with several representative baselines proposed for learning on imbalanced/skewed datasets in Table~\ref{table:results_main}. For a fair comparison, all the methods use the same two-tower architecture described in Section~\ref{sec:preliminaries}. We first observe that the baseline \ac{MUR} has marginal gains compared to the vanilla \ac{SUR} in all the benchmarks. This is likely because of highly skewed long-tail item distributions in the real world datasets (See item distribution in Figure~\ref{fig:item_distribution_real}). This observation is consistent with our findings from synthetic datasets having high skewness (Figure ~\ref{fig:syn-avg-clusters}). Whereas, with \ac{IDW} we see significant gains in performance for all real-world benchmarks we considered. In fact, the relative gains with \ac{IDW}-based \ac{MUR} ranged from 1.5x - 10x compared to the gains observed with the naive MUR. Our results also clearly show that \ac{IDW} outperforms other imbalanced baselines that are typically used to address data imbalance. Finally, \ac{MUR} coupled with IDW yield the best overall performance, highlighting that IDW is able to successfully leverage the improved clustering structure of \ac{MUR} over \ac{SUR} to yield further improvements.

\subsubsection{Head vs. Tail Performance} We conduct a performance analysis on the head and tail items separately to get insights where the overall performance gains are coming from. We compare our method with the top three baseline methods from Table~\ref{table:results_main}; namely, \ac{MUR}, Qmargin, and Item Norm. The NDCG@20 metric for all the three benchmarks is shown in Figure~\ref{fig:head_tail_split_results}\footnote{ We omit HR@20 metric for this analysis as it follows the same pattern as NDCG@20.}.
Compared to the naive \ac{MUR}, we observe that \ac{IDW}-based \ac{MUR} consistently improves the performance on the tail items, while having better or comparable performance across the head items, leading to better overall performance. Whereas, both Qmargin and Item Norm focus primarily on the tail items, while sacrificing significant performance for the head items. This makes \ac{IDW} a much more appealing method compared to other methods proposed for training on imbalanced datasets.

\subsection{Ablation Study} 

\subsubsection{Effect of Number of User Representations}
\label{sec:ablation-M}
\begin{figure}[!h]
    \centering
    \begin{subfigure}[b]{0.44\linewidth}
      \centering
      \includegraphics[width=\linewidth, height=2.7cm]{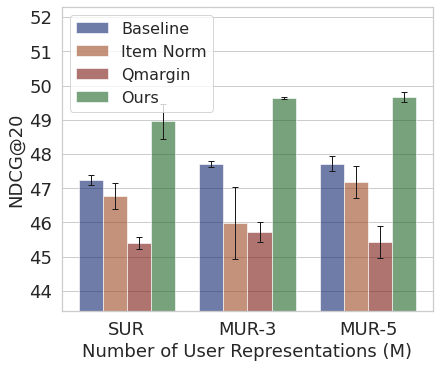} 
     \label{fig:ablation_heads_ndcg20_movielens}
    \end{subfigure}
    \hfill
    \begin{subfigure}[b]{0.44\linewidth}
      \centering
      \includegraphics[width=\linewidth, height=2.7cm]{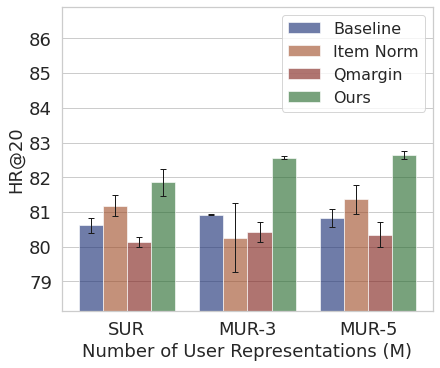} 
     \label{fig:ablation_heads_hr20_movielens}
    \end{subfigure}
    \vspace{-2em}
    \caption{The effect of number of user representations (M) on all the methods for the MovieLens-1M dataset.}
    \label{fig:ablations_M}
\end{figure}

We analyze the effect of the number of user representations ($M$) on the model performance for several representative methods. As shown in figure~\ref{fig:ablations_M}, \ac{IDW}-based models not only perform better than other counterparts for the same $M$, they are the only ones that show gains with increasing $M$. In contrast, the baseline methods do not observe any trend with respect to $M$. This shows that MUR and \ac{IDW} nicely complement each other.
\begin{figure}[!t]
    \centering
    \begin{subfigure}[b]{0.42\linewidth}
      \centering
      \includegraphics[width=\linewidth]{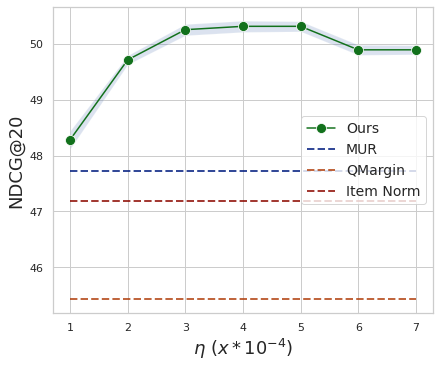}
      \caption{Ablation with varying $\eta$.}
     \label{fig:ablation_eta_ndcg20_movielens}
    \end{subfigure}
    \hfill
    \begin{subfigure}[b]{0.42\linewidth}
      \centering
      \includegraphics[width=\linewidth]{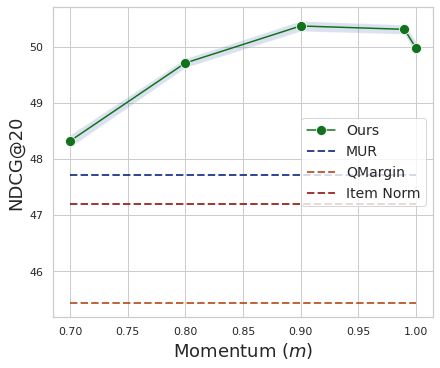}
      \caption{Ablation with varying $m$.}
     \label{fig:ablation_momentum_ndcg20_movielens}
    \end{subfigure}
    \vspace{-1em}
    \caption{The effect of $\eta$ and momentum (m) on \ac{IDW} performance for the MovieLens-1M dataset.}
    \label{fig:ablations_idw_hyperparams}
\end{figure}
\begin{figure}[!ht]
    \centering
    \begin{subfigure}[b]{0.47\linewidth}
      \centering
      \includegraphics[width=\linewidth, height=2.9cm]{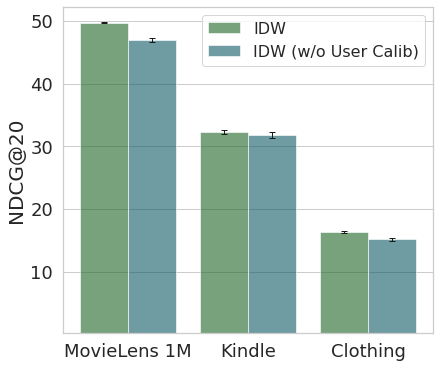} 
     \label{fig:ablation_calib_ndcg20}
    \end{subfigure}
    \hfill
    \begin{subfigure}[b]{0.47\linewidth}
      \centering
      \includegraphics[width=\linewidth, height=2.9cm]{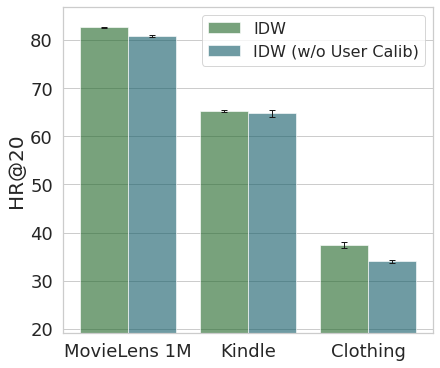} 
     \label{fig:ablation_calib_hr20}
    \end{subfigure}
    \vspace{-2em}
    \caption{The effect of user tower calibration for all datasets.}
    \label{fig:ablations_calib}
    \vspace{-1em}
\end{figure}
\vspace{-1em}
\subsubsection{Sensitivity to IDW Hyperparameters}
\label{sec:ablation-idw-hyperparams}
We also analyze the effect of the two hyperparameters introduced with \ac{IDW}: $\eta$ and momentum ($m$). The hyperparameter $\eta$ is used to define the stopping criterion in \ac{IDW} (See Algorithm~\ref{alg:idw}). Figure~\ref{fig:ablation_eta_ndcg20_movielens} shows the performance for $\eta \in [0.0001, 0.0007]$. The model performance can be sensitive to $\eta$; a value too low increases the number of iterations causing over-fitting, whereas a value too high reduces the number of iterations causing under-fitting. We found $\eta=0.0004$ to be the best in this ablation. We also perform an ablation for $m$, which determines the update in item weights (Equation~\ref{eq:update_weights}). We found that setting a low value for $m$ makes the model unstable due to an abrupt change in the loss function. In our ablation, $m=0.9$ yields the best results, followed by $m=0.99$ and $m=1.0$. The setting with $m=1$ doesn't use density-based weights and is equivalent to the baseline MUR with our proposed user tower calibration.

\subsubsection{Importance of User Tower Calibration}
\label{sec:ablation-idw-calibration}
In section~\ref{section:user-tower-calibration}, we proposed calibrating the user tower on the unweighted training distribution with the item tower fixed, once the model is trained with \ac{IDW}. To evaluate the effectiveness of calibrating the user tower, we do an ablation study comparing \ac{IDW} with and without user tower calibration (\ac{IDW} w/o user calib). Figure~\ref{fig:ablations_calib} shows that the proposed calibration of user tower boosts the overall recommendation performance, making it  an essential component of \ac{IDW}.

\section{Conclusions}

In conclusion, this paper makes contributions in two key aspects. Firstly, to the best of our knowledge, this is the first study to investigate the impact of skewed item distribution on learning \ac{MUR}, emphasizing the importance of loss balancing to fully realize the benefits of \ac{MUR}. This finding alone could hold substantial interest for the user-modeling community. Secondly, the paper introduces a novel density-weighting framework that utilizes item representations to balance the loss, resulting in improved performance with \ac{MUR}. Experimental results demonstrate that \ac{MUR} with \ac{IDW} yields superior overall performance on both public benchmarks and synthetic datasets, highlighting that \ac{IDW} is successfully able to use improved learned item representation to improve overall recommendation performance. The presented work also sets the stage for further exploration in the field, opening avenues for the development of alternative balancing techniques and more robust \ac{MUR} extraction layers. We hope that the findings in this work contribute to advancing the field of user modeling in recommendation systems and pave the way for future research endeavors in this domain.

\appendix
\appendixhead
\vspace{-1em}
\section{Synthetic Data Generation}
\label{apx:synthetic_data}
\noindent We generate synthetic data containing a sequence of item clicks for each user. The generated item sequences are used for training on the next item prediction task. The parameters for generating the dataset are summarized in Table~\ref{tab:parameters_synthetic}. The data generation process is specified as follows:
\begin{enumerate}[leftmargin=1.5em]
    \item \textbf{Assign clusters to users:} We assign each user a set of clusters ($Y_u$) that depict user's interests. The cluster assignment is done by sampling from a multinomial distribution given by weights $\{\pi_1,\dots,\pi_C\}$. 
    The sampling weights are chosen to follow a power-law, which results in skewness in the user interest distribution as shown in Figure~\ref{fig:syn-cluster-distrib}.
    
    \item \textbf{Generate cluster sequence:} Given a user's interests, we assume a finite state Hidden Markov model to define the cluster transitions over time. The resulting sequence reflects a user's interest over time. The cluster transition is given by:
    \begin{equation}
    \label{eq:synthetic_cluster_transition}
    P^{(u)}_{ij} = \begin{cases}
         \alpha & \text{if } i=j \text{ and } i\in Y_u\\
         \frac{\gamma}{|Y_u - 1|} & \text{if } i \neq j \text{ and } \{i,j\} \in Y_u\\[4pt]
         \frac{1 - \alpha - \gamma}{C - |Y_u|} & \text{if } j \notin Y_u \text{ and } i\in Y_u\\
         \epsilon & \text{if } i=j \text{ and } i\notin Y_u\\
         \frac{1 - \epsilon}{|Y_u|} & \text{if } j \in Y_u \text{ and } i\notin Y_u\\
         0 & \text{if } i \neq j \text{ and } \{i, j\} \notin Y_u
    \end{cases}
\end{equation}
    where $P^{(u)}_{ij}$ depicts the probability of transition from cluster $i$ to cluster $j$ for a user $u$. In (\ref{eq:synthetic_cluster_transition}), $\gamma$ depicts the user's interest volatility: a high $\gamma$ would lead to frequent switching between different clusters that the user is interested in.
    \item \textbf{Generate item sequence:} Finally, we generate the item sequence using the cluster sequence generated above. We assume that the items from each cluster are drawn based on a multinomial distribution with $\{p_1,\dots,p_{K}\}$, where $K$ is the number of items. For all clusters, we assume $K = |\gI|/C$. Similar to cluster assignment, we induce a power-law distribution for $\{p_1,\dots,p_{K}\}$ as shown in Figure~\ref{fig:syn-cond-item-distrib}.
\end{enumerate}
\begin{table}[!ht]
    \centering
    \caption{Parameters used to generate the synthetic sequential item history.}
    \label{tab:parameters_synthetic}
    \resizebox{0.35\textwidth}{!}{
    \begin{tabular}{c|p{0.75\linewidth} }
        \toprule
        Param & Description \\
        \midrule
        $C$ &  The number of total clusters or interests.\\
        $T$ &  The sequence length.\\
        $|\gU|$ & The number of total users. \\
        $|\gI|$ & The number of total items. \\
        $Y_u$ & The set of clusters that a user $u$ is interested in. \\
        $\alpha$ & Probability of staying in the current cluster $c\in Y_u$. \\
        $\gamma$ & User interest volatility: The total probability of a user switching to a new cluster in $Y_u$.\\
        $\epsilon$ & Probability that a user stays in a ``non-interesting" cluster $c \notin Y_u$. \\
        \bottomrule
    \end{tabular}}
\end{table}
Note that the ground-truth cluster correspondences are not available to the recommendation model during training on the synthetic dataset, $i.e.$ the model only has access to the marginal item distribution (Figure~\ref{fig:syn-item-distrib}). \\ 

\noindent\textbf{Parameters for generating synthetic data}: For the experiments presented in Section~\ref{section:synthetic-analysis}, we use a set of 50 items with 10 items per interest (cluster). The degree of skewness in Figure~\ref{fig:syn-results} is set by changing the imbalance in the cluster sampling distribution governed by $\pi_c$: the power law exponent is increased from 0.0 to 2.0 to make the interest distribution skewed. The conditional item sampling distribution $p_k$ also uses power law weights, however, for all the experiments the item sampling distribution uses 0.5 as the power law exponent. For all experiments, we set the interest volatility parameters $\gamma = 0.3$ and $\alpha=0.6$. We set $\epsilon=0.1$ to induce exploration in the user engagements.

\input{Tables/imbalance_overall_k5_single_column}

\section{Recommendation Performance measured as HR@5 and NDCG@5}
\label{apx:results_k5}
In addition to the main results showing the recommendation performance measured as HR@20 and NDCG@20 in Table~\ref{table:results_main}, we also include metrics HR@5 and NDCG@5 for overall item sets in Table~\ref{table:results_main_k5}. Once again, we observe that \ac{IDW} performs better than the other baselines we considered.

\section{Item distribution in the datasets}
\label{apx:item_distr}
\begin{figure}[!ht]
    \centering
    \includegraphics[width=\linewidth]{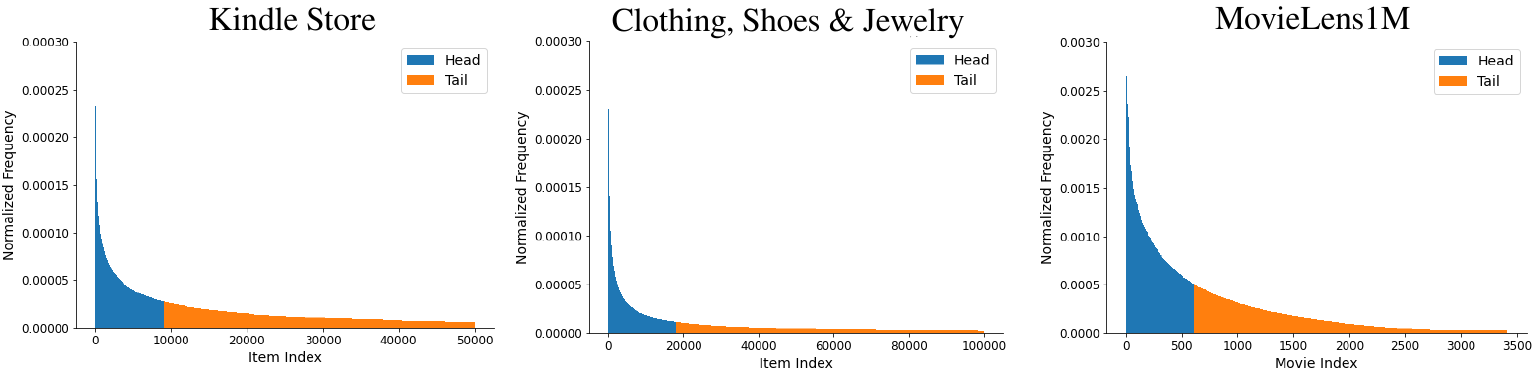}
    \caption{The item distribution in the three benchmarks. (Left) Kindle Store. (Middle) Clothing, Shoes and Jewelry. (Right) MovieLens-1M.}
    \label{fig:item_distribution_real}
\end{figure}

In this section, we visualize the empirical distribution of the three benchmarks that we considered in our experiments in Figure~\ref{fig:item_distribution_real}. In the y-axis we plot the normalized frequency of items and x-axis is the item ids arranged based on their frequency.

\clearpage
\bibliographystyle{ACM-Reference-Format}
\bibliography{references}

\begin{acronym}[CCCCCCCC]
    \acro{DNN}{Deep Neural Network}
    \acro{DNNs}{Deep Neural Networks}
    \acro{MS}{mean-silhouette}
    \acro{IDW}{Iterative Density Weighting}
    \acro{KDE}{Kernel Density Estimation}
    \acro{MUR}{multiple user representations}
    \acro{SUR}{single user representation}
\end{acronym}

\end{document}

%% file: math_commands.tex
\usepackage{color}
\usepackage{amsmath,amsthm,amssymb}
\usepackage[mathscr]{euscript}

\def\eqref#1{equation~\ref{#1}}

\def\1{\bm{1}}

\DeclareMathAlphabet{\mathsfit}{\encodingdefault}{\sfdefault}{m}{sl}
\SetMathAlphabet{\mathsfit}{bold}{\encodingdefault}{\sfdefault}{bx}{n}

\def\gB{{\mathcal{B}}}

\def\gD{{\mathcal{D}}}

\def\gI{{\mathcal{I}}}

\def\gK{{\mathcal{K}}}
\def\gL{{\mathcal{L}}}

\def\gU{{\mathcal{U}}}

\newcommand{\E}{\mathbb{E}}

\DeclareMathOperator*{\argmin}{arg\,min}

%% file: Tables/imbalance_overall_k20.tex
\begin{table*}[!ht]
\centering
\caption{The recommendation performance on three real-world benchmark datasets. We use blue color for the best baseline. We use bold and underline to denote the best metric. (Higher is better).}
\label{table:results_main}
\resizebox{0.9\textwidth}{!}{
\begin{tabular}{lc@{\hskip 1em}*{6}{c}}
\toprule
Method & \# User Representations (M) & \multicolumn{2}{c}{\textbf{MovieLens 1M}} & \multicolumn{2}{c}{\textbf{Kindle Store}} &  \multicolumn{2}{c}{\textbf{Clothing, Shoes and Jewelry}} \\
\arrayrulecolor{gray}\cmidrule(r){3-4}\cmidrule(r){5-6}\cmidrule(r){7-8}
&  & HR@20 & NDCG@20 & HR@20 & NDCG@20 & HR@20 & NDCG@20 \\
\midrule
SUR~\cite{Humeau2020PolyEncoders} & 1& 80.62 $\pm$ 0.18 & 47.24 $\pm$ 0.14 & 63.33 $\pm$ 0.38 & 30.87 $\pm$ 0.56 & 32.43 $\pm$ 0.66 & 13.44 $\pm$ 0.54\\
MUR~\cite{Humeau2020PolyEncoders} & 5 & 80.82 $\pm$ 0.20 & \second{47.72} $\pm$ 0.23 & 64.66 $\pm$ 0.45 & 31.16 $\pm$ 0.38 & 33.92 $\pm$ 0.94 & \second{14.90} $\pm$ 0.52 \\
\arrayrulecolor{gray}\cmidrule(r){1-8}
Frequency Balance~\cite{he2009Imbalance} & 5 &  75.75 $\pm$ 0.32 & 40.36 $\pm$ 0.24 & 55.96 $\pm$ 0.77 & 23.49 $\pm$ 0.33 & 22.38 $\pm$ 0.52 & 7.44 $\pm$ 0.31 \\
Focal Balance~\cite{Lin2018FocalLoss} & 5 & 80.88 $\pm$ 0.22 & 45.48 $\pm$ 0.37 & 62.17 $\pm$ 0.94 & 28.56 $\pm$ 0.83 & 32.39 $\pm$ 1.16 & 13.19 $\pm$ 0.61\\
Qmargin~\cite{CaoMargin2019} & 5 & 80.35 $\pm$ 0.37 & 45.43 $\pm$ 0.46 & 63.75 $\pm$ 0.71 & 29.87 $\pm$ 0.50 & \second{34.71} $\pm$ 1.16 & 14.13 $\pm$ 0.52\\
Item Norm ~\cite{kim2020Imbalance} & 5 & \second{81.36} $\pm$ 0.42 & 47.19 $\pm$ 0.47 & \second{65.20} $\pm$ 0.49 & \second{32.19} $\pm$ 0.24 & 34.57 $\pm$ 0.67 & 14.59 $\pm$ 0.71 \\
Item Norm (Post-hoc)~\cite{Kang2020Decoupling} & 5 & 80.20 $\pm$ 0.53 & 45.87 $\pm$ 0.67 & 64.05 $\pm$ 0.18 & 30.68 $\pm$ 0.22 & 34.32 $\pm$ 0.77 & 13.83 $\pm$ 0.52\\
\arrayrulecolor{gray}\cmidrule(r){1-8}
\textbf{SUR-IDW [Ours]} & 1 & 81.85 $\pm$ 0.40 & 48.95 $\pm$ 0.51 & 64.62 $\pm$ 0.28 & 31.67 $\pm$ 0.26 & 34.88 $\pm$ 0.38 & 15.45 $\pm$ 0.16 \\
\textbf{MUR-IDW [Ours]} & 5 &\underline{\textbf{82.65}} $\pm$ 0.03 & \underline{\textbf{49.67}} $\pm$ 0.14 & \underline{\textbf{65.24}} $\pm$ 0.25 & \underline{\textbf{32.25}} $\pm$ 0.32 & \underline{\textbf{37.34}} $\pm$ 0.63 & \underline{\textbf{16.33}} $\pm$ 0.18 \\
\arrayrulecolor{gray}\cmidrule(r){1-8}
MUR vs. SUR & - & \green{+0.25\%} & \green{+1.02\%} & \green{+2.10\%} & \green{+0.94\%} & \green{+4.59\%} & \green{+10.86\%}\\
\textbf{MUR-IDW [Ours]} vs. SUR & - & \green{+2.52\%} & \green{+5.14\%} & \green{+3.02\%} & \green{+4.47\%} & \green{+15.14\%} & \green{+21.50\%}\\
\bottomrule
\end{tabular}
}
\end{table*}

%% file: Tables/dataset_summary.tex
\begin{table}[!ht]
\centering
\caption{Dataset Statistics.}
\vspace{-1em}
\label{table:dataset_summary}
\resizebox{\columnwidth}{!}{%
\begin{tabular}{lllll}
\toprule
Dataset & \# Users &  \# Items & Avg Seq Length
\\
\midrule
MovieLens-1M~\cite{harper2015MovieLens} & 6,040 & 3,706 & 165\\
Kindle Store~\cite{ni2019justifying} & 139,769 & 98,769 & 15\\
Clothing Shoes \& Jewelry~\cite{ni2019justifying} & 1,219,594 & 376,438 & 9\\
\bottomrule
\vspace{-2em}
\end{tabular}}
\end{table}

%% file: Tables/imbalance_overall_k5_single_column.tex
\begin{table}[!ht]
\centering
\caption{The recommendation performance on three real-world benchmark datasets with K=5. }
\vspace{-1em}
\label{table:results_main_k5}
\resizebox{0.5\textwidth}{!}{
\begin{tabular}{lc@{\hskip 1em}*{5}{c}}
\toprule
Method 
& \multicolumn{2}{c}{\textbf{MovieLens 1M}} & \multicolumn{2}{c}{\textbf{Kindle Store}} &  \multicolumn{2}{c}{\textbf{Clothing, Shoes ...}} \\
\arrayrulecolor{gray}\cmidrule(r){2-3}\cmidrule(r){4-5}\cmidrule(r){6-7}
&  HR@5 & NDCG@5 & HR@5 & NDCG@5 & HR@5 & NDCG@5 \\
\midrule
SUR 
& 54.69 $\pm$ 0.20 & 39.72 $\pm$ 0.14 & 32.98 $\pm$ 0.85 & 22.13 $\pm$ 0.66 & 11.87 $\pm$ 0.67 & 7.76 $\pm$ 0.55\\
MUR 
& \second{55.21} $\pm$ 0.29 & \second{40.25} $\pm$ 0.24 & 33.44 $\pm$ 0.58 & 22.35 $\pm$ 0.44 & \second{13.13} $\pm$ 0.63 & \second{9.15} $\pm$ 0.43\\
\arrayrulecolor{gray}\cmidrule(r){1-7}
Frequency Bal. 
& 46.07 $\pm$ 0.34 & 31.72 $\pm$ 0.25 & 23.10 $\pm$ 0.31 & 14.46 $\pm$ 0.22 &  5.34 $\pm$ 0.31 & 3.16 $\pm$ 0.19\\
Focal Bal. 
& 53.27 $\pm$ 0.48 & 37.42 $\pm$ 0.43 &  30.93 $\pm$ 1.21 & 19.65 $\pm$ 0.90 & 11.49 $\pm$ 0.73 & 7.42 $\pm$ 0.49\\
Qmargin 
& 52.80 $\pm$ 0.63 & 37.40 $\pm$ 0.53 & 32.03 $\pm$ 0.64 & 20.82 $\pm$ 0.47 & 12.32 $\pm$ 0.53 &  7.94 $\pm$ 0.42\\
Item Norm 
& 54.93 $\pm$ 0.67 & 39.48 $\pm$ 0.54 & \second{35.02} $\pm$ 0.44 &  \second{23.55} $\pm$ 0.37 & 12.68 $\pm$ 0.79 &  8.55 $\pm$ 0.75\\
Post-hoc 
& 53.09 $\pm$ 0.92 & 37.98 $\pm$ 0.77 & 32.38 $\pm$ 0.37 & 21.66 $\pm$ 0.32 & 11.94 $\pm$ 0.61 & 7.64 $\pm$ 0.50\\
\arrayrulecolor{gray}\cmidrule(r){1-7}
\textbf{SUR-IDW} 
&  56.88 $\pm$ 0.69 & 41.57 $\pm$ 0.59 & 34.29 $\pm$ 0.31 & 23.01 $\pm$ 0.28 & 13.79 $\pm$ 0.18 & 9.62 $\pm$ 0.10\\
\textbf{MUR-IDW} 
& \underline{\textbf{57.79}} $\pm$ 0.16 & \underline{\textbf{42.42}} $\pm$ 0.15 & \underline{\textbf{35.07}} $\pm$ 0.51 & \underline{\textbf{23.62}} $\pm$ 0.39 & \underline{\textbf{14.67}} $\pm$ 0.16 & \underline{\textbf{10.43}} $\pm$ 0.20\\
\bottomrule
\end{tabular}
}
\end{table}